\documentclass[12pt]{iopart}
\usepackage{graphicx}  
  
\begin{document}

\title[Photoionization in correlated three-atomic systems]{Resonantly enhanced photoionization in correlated three-atomic systems}

\author{B Najjari$^{1}$, C M\"uller$^{1,2}$ and A B Voitkiv$^{1}$}

\address{$^{1}$ Max-Planck-Institut f\"ur Kernphysik, Saupfercheckweg 1, 69117 Heidelberg, Germany}
\address{$^{2}$ Institut f\"ur Theoretische Physik I, Heinrich-Heine Universit\"at D\"usseldorf, Universit\"atsstra{\ss}e 1, 40225 D\"usseldorf, Germany}
\ead{najjari@mpi-hd.mpg.de, c.mueller@tp1.uni-duesseldorf.de, voitkiv@mpi-hd.mpg.de}

\begin{abstract}
Modifications of photoionization arising from resonant electron-electron
correlations between neighbouring atoms in an atomic sample are studied.
The sample contains atomic species $A$ and $B$, with the ionization 
potential of $A$ being smaller than the energy of a dipole-allowed transition in $B$. 
The atoms are subject to an external radiation field which 
is near-resonant with the dipole transition in $B$. Photoionization of an atom $A$ may thus proceed via a
two-step mechanism: photoexcitation in the subsystem of species $B$, followed by interatomic
Coulombic decay. As a basic atomic configuration, we investigate resonant 
photoionization in a three-atomic system $A$-$B$-$B$, consisting of an atom $A$ and two neighbouring atoms $B$. 
It is found that, under suitable conditions, the influence of the neighbouring
atoms can strongly affect the photoionization process, including its total
probabilty, time development and photoelectron spectra. In particular, by comparing our 
results with those for photoionization of an isolated atom $A$ and a two-atomic system $A$-$B$, 
respectively, we reveal the characteristic impact exerted by the third atom.

\end{abstract}


\section{Introduction}
Photoionization of atoms and molecules is one of the most fundamental quantum processes. It played a key role in the early days of quantum mechanics and has ever since been paving the way towards an improved understanding of the structure and dynamics of matter on a microscopic scale. Today, kinematically complete photoionization experiments allow for accurate tests of the most sophisticated ab-initio calculations. Besides, photoionization studies in a new frequency domain are currently becoming feasible by the availability of novel xuv and x-ray radiation sources \cite{FLASH,LCLS,atto}, giving rise to corresponding theoretical developments (see, e.g., \cite{Lambropoulos,Dieter,Santra}).

Various photoionization mechanisms rely crucially on electron-electron correlations. Prominent examples are single-photon double ionization as well as resonant photoionization. The latter proceeds through resonant photoexcitation of an autoionizing state with subsequent Auger decay. In recent years, a similar kind of ionization process has been studied in systems consisting of two (or more) atoms. Here, a resonantly excited atom transfers its excitation energy radiationlessly via interatomic electron-electron correlations to a neighbouring atom leading to its ionization. This Auger-like decay involving two atomic centers is commonly known as interatomic Coulombic decay (ICD) \cite{ICD,ICDrev}. It has been observed, for instance, in noble gas dimers and water molecules \cite{ICDexp}. In metal oxides, the closely related process of multi-atom resonant photoemission (MARPE) was also observed \cite{MARPE}.

We have recently studied resonant two-center photoionization in heteroatomic systems and shown that this ionization channel can be remarkably strong \cite{2CPI,ABV1,ABV2}. In particular, it can dominate over the usual single-center photoionization by orders of magnitude.  Besides, characteristic effects resulting from a strong coupling of the ground and autoionizing states by a relatively intense photon field were identified. Also resonant two-photon ionization in a system of two identical atoms was investigated \cite{identical}. We note that photoionization in two-atomic systems was also studied in \cite{Kuhn,Kuhn2} and \cite{Perina,Perina2}. The inverse of two-center photoionization (in weak external fields) is two-center dielectronic recombination \cite{2CDR}.

\begin{figure}[b]  
\begin{center}
\includegraphics[width=0.45\textwidth]{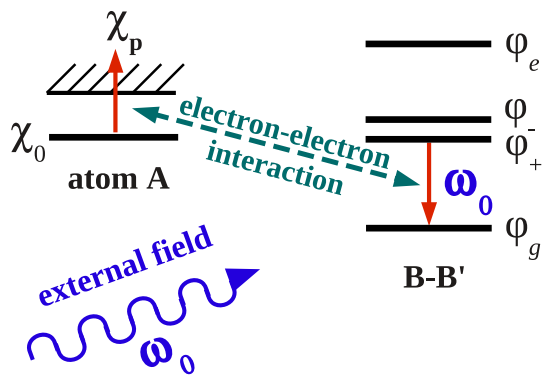}
\end{center}
\caption{Schematic illustration of photoionization of an atom $A$ in the presence of an external laser field and two neighbouring atoms $B$ and $B'$. Apart from the direct photoionization of $A$ there are interatomic channels via resonant photoexcitation of the ``molecular'' system $B$-$B'$ and subsequent ICD. }
\label{figure1}
\end{figure}

In the present contribution, we extend our investigations of electron correlation-driven interatomic processes by considering photoionization of an atom $A$ in the presence of \textit{two} neighbouring atoms $B$ (see figure 1). All atoms are assumed to interact with each other and with an external radiation field. We show that the photoionization of atom $A$ via photoexcitation of the system of two neighbouring atoms $B$ and subsequent ICD can be by far the dominant ionization channel. Moreover, we reveal the characteristic properties of the process with regard to its temporal dependence and photoelectron spectra. In particular, by comparing our results with those for photoionization in a system of two atoms $A$ and $B$, we demonstrate the influence which the presence of the second atom $B$ may have.

Atomic units (a.u.) are used throughout unless otherwise stated.

\section{Theoretical Framework}

Let us consider a system consisting
of three atoms, $A$, $B$ and $B'$, where $B$ and $B'$ are
atoms of the same element and $A$ is different.
We shall assume that all these atoms are separated
by sufficiently large distances such that free atomic states
represent a reasonable initial basis set to start with.

Let the ionization potential $ I_A $
of atom $A$ be smaller than
the excitation energy $ \Delta E_B $
of a dipole-allowed transition in atoms $B$ and $B'$.
Under such conditions, if our system is irradiated by
an electromagnetic field with frequency $\omega_0 \approx \Delta E_B$, the
ionization process of this system
(i.e., essentially of the atom $A$) can be qualitatively
different compared to the case when a single, isolated atom
$A$ is ionized. Indeed, in such a case $A$ can be ionized
not only directly but also via resonant photoexcitation
of the subsystem of $B$ and $B'$, with its consequent deexcitation
through energy transfer
to $A$ resulting in ionization of the latter.

In the following, we consider
photoionization in the system of atoms $A$, $B$ and $B'$ in more detail.
For simplicity, we suppose that the nuclei
of all atoms are at rest during photoionization.
Denoting the origin of our coordinate system by $O$,
we assume that the nuclei of the atoms $B$ and $B'$
are located on the $Z$-axis: ${\bf R}_B = (0,0,Z_B)$ and
${\bf R}_{B'} = (0,0,Z_{B'})$. The coordinates of the nucleus of
the atom $A$ are given by ${\bf R}_A = (X_A,Y_A,Z_A)$.
The coordinates of the (active) electron of
atom $\lambda$ with respect to its nucleus are denoted by
${\bf r}_{\lambda}$, where $\lambda \in \{A, B, B'\}$.

The total Hamiltonian describing the
three atoms embedded in an external electromagnetic field reads
\begin{eqnarray}
H =  \hat{H}_0 + \hat{V}_{AB} + \hat{V}_{AB'} + \hat{V}_{BB'} + \hat{W}_A
+ \hat{W}_B + \hat{W}_{B'},
\label{hamiltonian}
\end{eqnarray}
where $ \hat{H}_0 $ is the sum of the Hamiltonians
for the noninteracting atoms $A$, $B$ and $B'$.

We shall assume that the (typical) distances
$\Delta R$ between the atoms
are not too large, $ \Delta R \ll c / \Delta E_B $,
where $c$ is the speed of light, such that retardation
effects in the electromagnetic interactions can be ignored.
If transitions of electrons between bound states
in atoms $B$ and $B'$ are of dipole character,
then the interaction between each pair of atoms
$(\lambda,\gamma)$ (with $\lambda, \gamma \in \{A, B, B'\}$)
can be written as
\begin{eqnarray}
\hat{V}_{\lambda,\gamma} &=&
\frac{ \left( {\bf r}_{\lambda} \right)_i \left( {\bf r}_{\gamma}
\right)_j }{R_{\lambda,\gamma}^3}
\left( \delta_{ij} - \frac{ 3 \left( {\bf R}_{\lambda,\gamma} \right)_i \left(
{\bf R}_{\lambda,\gamma} \right)_j }
{R_{\lambda,\gamma}^2} \right),
\label{inter_atomic}
\end{eqnarray}
where $\bf{R}_{\lambda,\gamma} = \bf{R}_{\lambda} - \bf{R}_{\gamma}$
and $\delta_{ij}$ is the Kronecker symbol.
Note that in (\ref{inter_atomic})
a summation over the repeated indices $i$ and $j$ is implied.

In (\ref{inter_atomic}), $\hat{W}_{\lambda}$ denotes
the interaction of the atom $\lambda$ with
the laser electromagnetic field. The latter will be treated
as a classical, linearly polarized field,
described by the vector potential
${\bf A}({\bf r},t)= {\bf A}_0 \cos\left(\omega_0 t \right)$,
where ${\bf A}_0 = c {\bf F}_0/\omega_0$,
$\omega_0 = c k_0 $ is the angular frequency
and ${\bf F}_0$ is the field strength.
The interaction $\hat{W}_{\lambda}$ then reads
\begin{eqnarray}
\hat{W}_{\lambda} = \frac{1}{c} {\bf A}({\bf r}_{\lambda},t) \cdot \hat{\bf
p}_{\lambda},
\label{interaction}
\end{eqnarray}
where $\hat{\bf p}_{\lambda}$ is the momentum operator for
the electron in atom ${\lambda}$.

Our treatment of photoionization will be based on
the following points:

Oscillator strengths for dipole-allowed
bound-bound transitions can be very strong.
This means that, provided that the distances between
all the atoms in our system are
of the same order of magnitude,
the interaction between atoms $B$ and $B'$
is much more effective than the interaction between atoms
$A$ and $B$ (or $A$ and $B'$). Besides, atoms $B$ and $B'$
will, in general, couple much more strongly to
a resonant laser field than atom $A$.
In what follows, we shall assume that
the intensity of the laser field is relatively low such
that the interaction between atoms $B$ and $B'$
changes the states of the system more substantially than
the coupling of these atoms to the laser field.
Therefore, we shall begin with building states of
the $B$-$B'$ subsystem in the absence of the field.
The second step of our treatment will be to
include the interaction of the $B$-$B'$
subsystem with the laser field and,
in the third step, we complete the treatment of ionization
by considering the interaction of atom $A$ with both the laser field
and the field-dressed subsystem of atoms $B$ and $B'$.

\vspace{1cm}

I. We denote the ground and excited states
of the undistorted atoms $B$ and $B'$ by
$\phi_0$, $\phi_e$ and $\phi'_0$, $\phi'_e$, respectively.
Let the corresponding energies of these states be $\varepsilon_0$ and
$\varepsilon_e$.
The state $\psi_{BB'}$ of the $B$-$B'$ subsystem can be expanded
into the ``complete'' set of undistorted atomic states
represented by the configurations
(i) $\phi_0 \phi'_0$, (ii) $\phi_0 \phi'_e$,
(iii) $\phi_e \phi'_0$ and (iv) $\phi_e \phi'_e$.
In the approximation, which neglects the interatomic interaction,
the configurations $\phi_0 \phi'_e$ and $\phi_e \phi'_0$
are characterized by exactly the same value of
the (undistorted) energy $E_{0e} = \varepsilon_0 + \varepsilon_e $.
The latter, in turn, strongly differs from
the energies $E_{00} = 2 \varepsilon_0 $ and
$E_{ee} = 2 \varepsilon_e $ which are
characteristic for the configurations
$\phi_0 \phi'_0$ and $\phi_e \phi'_e$,
respectively. Therefore, provided that the distance
between the atoms is not too small,
the interaction $V_{BB'}$
will strongly mix the configurations (ii) and (iii) only,
while the other configurations (i) and (iv)
will be affected only very weakly.
Taking this into account, it is not difficult
to find the states of the subsystem of
interacting atoms $B$ and $B'$ which read
\begin{eqnarray}
\varphi_{g} &=&  \phi_0 \phi'_0
\nonumber \\
\varphi_{+}  &=& \frac{1}{\sqrt{2}}
\left( \phi_e \phi'_0 + \phi_0 \phi'_e \right)
\nonumber \\
\varphi_{-}  &=& \frac{1}{\sqrt{2}}
\left( \phi_e \phi'_0 - \phi_0 \phi'_e \right)
\nonumber \\
\varphi_{e} &=& \phi_e \phi'_e.
\label{states_of_BB'}
\end{eqnarray}
These two-atomic states are normalized
and mutually orthogonal. They posses energies given by
$E_{g} = 2 \varepsilon_0 $,
$ E_+ = \varepsilon_0 + \varepsilon_e + v_{BB'} $,
$ E_- = \varepsilon_0 + \varepsilon_e - v_{BB'} $
and $E_{e} = 2 \varepsilon_e $, respectively, where
$ v_{BB'} = \left\langle \phi_e \phi'_0 \left| \hat{V}_{BB'} \right|
\phi_0 \phi'_e  \right\rangle$. Note that, for definiteness, 
$ v_{BB'}$ has been assumed to be real and negative here, 
as will always be the case in our examples below (see
section~3).

\vspace{1cm}

II. Let us now consider two interacting atoms $B$ and $B'$
embedded in a resonant laser field. One can look for a state
of such a system by expanding it into the new set of states given by
Eq.~(\ref{states_of_BB'}),
\begin{eqnarray}
\psi(t) &=& g(t)\varphi_{g} + a_{+}(t) \varphi_{+} +
a_{-}(t) \varphi_{-} +  b(t) \varphi_{e}.
\label{BB'_in_laser_1}
\end{eqnarray}
Inserting the expansion (\ref{BB'_in_laser_1})
into the corresponding wave equation, we obtain
a set of coupled equations for the unknown time-dependent coefficients
$g(t)$, $a_{+}(t)$, $a_{-}(t)$ and $b(t)$:
\begin{eqnarray}
i \frac{dg}{dt} - E_g g &=&
\left\langle \varphi_g \left|\hat{W}_{B} + \hat{W}_{B'}\right| \varphi_{+}
\right\rangle a_{+} +
\left\langle \varphi_g \left|\hat{W}_{B} + \hat{W}_{B'}\right| \varphi_{-}
\right\rangle a_{-} \nonumber\\
& &  +
\left\langle \varphi_g \left|\hat{W}_{B} + \hat{W}_{B'}\right| \varphi_{e}
\right\rangle b
\nonumber \\
i \frac{da_{+}}{dt} - E_{+} a_{+} &=&
\left\langle \varphi_{+} \left|\hat{W}_{B} + \hat{W}_{B'}\right| \varphi_g
\right\rangle g +
\left\langle \varphi_{+} \left|\hat{W}_{B} + \hat{W}_{B'}\right|
\varphi_{-} \right\rangle a_{-} \nonumber\\
& & +
\left\langle \varphi_{+} \left|\hat{W}_{B} + \hat{W}_{B'}\right|
\varphi_{e} \right\rangle b
\nonumber \\
i \frac{da_{-}}{dt} - E_{-}a_{-} &=&
\left\langle \varphi_{-} \left|\hat{W}_{B} + \hat{W}_{B'}\right| \varphi_g
\right\rangle g +
\left\langle \varphi_{-} \left|\hat{W}_{B} + \hat{W}_{B'}\right|
\varphi_{+} \right\rangle a_{+} \nonumber\\
& & +
\left\langle \varphi_{-} \left|\hat{W}_{B} + \hat{W}_{B'}\right|
\varphi_{e} \right\rangle b
\nonumber \\
i \frac{db}{dt} - E_e b &=&
\left\langle \varphi_e \left|\hat{W}_{B} + \hat{W}_{B'}\right| \varphi_g
\right\rangle g +
\left\langle \varphi_e \left|\hat{W}_{B} + \hat{W}_{B'}\right| \varphi_{+}
\right\rangle a_{+} \nonumber\\
& & +
\left\langle \varphi_e \left|\hat{W}_{B} + \hat{W}_{B'}\right| \varphi_{-}
\right\rangle a_{-}.
\label{BB'_in_laser_2}
\end{eqnarray}
The system of equations (\ref{BB'_in_laser_2}) can be greatly simplified
by noting the following. First, all transition matrix
elements of the interaction with the laser field, which involve the
asymmetric
state $\varphi_{-}$, are equal to zero and, thus, only the remaining
three states can be coupled by the field. Second, if we suppose that
the frequency of the laser field is resonant to the transitions
$ \varphi_g \longleftrightarrow \varphi_{+} $ and that the field is
relatively weak
such that the non-resonant transitions $ \varphi_{+} \longleftrightarrow
\varphi_e $
are much less effective than the above resonant ones, the system
(\ref{BB'_in_laser_2}) effectively reduces to
\begin{eqnarray}
i \frac{dg}{dt} - E_g g &=&
\left\langle \varphi_g \left|\hat{W}_{B} + \hat{W}_{B'}\right| \varphi_{+}
\right\rangle a_{+}
\nonumber \\
i \frac{da_{+}}{dt} - E_{+} a_{+} &=&
\left\langle \varphi_{+} \left|\hat{W}_{B} + \hat{W}_{B'}\right| \varphi_g
\right\rangle g,
\label{BB'_in_laser_3}
\end{eqnarray}
which can be readily solved by using the rotating wave approximation.
Assuming that the field is switched on suddenly at $t=0$,
we obtain two solutions
\begin{eqnarray}
\psi_1(t) &=& \frac{1}{z_{2} - z_{1}}
\left[ \left( z_{2} + \omega_0 - E_{+} \right)
{\rm e}^{-i z_{2} t} - \left( z_{1} + \omega_0 - E_{+} \right)
{\rm e}^{-i z_{1} t} \right] \, \varphi_{g}
\nonumber \\
&+& \frac{W_{+,g}}{z_{2} - z_{1}}
\left( {\rm e}^{-i z_{2} t} - {\rm e}^{-i z_{1} t} \right)
{\rm e}^{- i \omega_0 t} \, \varphi_{+}
\label{BB'_in_laser_4}
\end{eqnarray}
and
\begin{eqnarray}
\psi_2(t) &=& \frac{ W_{g,+} }{ z_{2} - z_{1} }
\left( {\rm e}^{-i z_{2} t} - {\rm e}^{-i z_{1} t} \right) \, \varphi_{g}
\nonumber \\
&+& \frac{1}{ z_{2} - z_{1} }
\left[ \left( z_{2} - E_{g} \right) {\rm e}^{-i z_{2} t}
- \left( z_{1} - E_{g} \right)
{\rm e}^{-i z_{1} t} \right] {\rm e}^{- i \omega_0 t} \, \varphi_{+}.
\label{BB'_in_laser_5}
\end{eqnarray}
In the above equations, we have introduced
\begin{eqnarray}
z_{1} &=& \frac{1}{2} %
\left( E_{g} + E_{+} - \omega_0 - \Omega_R \right)
\nonumber \\
z_{2} &=& \frac{1}{2} %
\left( E_{g} + E_{+} - \omega_0 + \Omega_R \right), %
\label{BB'_in_laser_6}
\end{eqnarray}
where $\Omega_R = \sqrt{( E_{+} - E_{g} - \omega_0)^2 + 4 \left| W_{g,+}
\right|^2} $
is the Rabi frequency, $W_{g,+} = \left\langle \varphi_g \left| %
{\bf F}_0 \cdot \left( \hat{\bf p}_B + \hat{\bf p}_{B'} \right)/%
(2 \omega_0) \right| \varphi_{+} \right\rangle $
and $W_{+,g}=(W_{g,+})^*$.

The two solutions in (\ref{BB'_in_laser_5})
correspond to two different initial conditions:
at $t=0$ the system is either in
the state $\varphi_{g}$ or in $\varphi_{+}$.
They are orthogonal to each other and
form a ``complete'' set of field-dressed states
of the subsystem $B$-$B'$.
Note also that we have neglected the spontaneous
radiative decay of the excited state $\varphi_{+}$ which,
in our case, is justified as long as $\left| W_{g,+} \right| \gg \Gamma_r$,
where $\Gamma_r$ is the radiative width of $\varphi_{+}$.

\vspace{1cm}

III. Now, as the last step, we shall add atom $A$ to our consideration.
Let $\chi_0$ and $\chi_{\bf p}$, where ${\bf p}$ is the electron momentum,
be the ground and a continuum state of a single, isolated atom $A$.
The wavefunction of the total system $A$--$B$-$B'$ can be expanded
into the following ``complete'' set of states
\begin{eqnarray}
\Psi(t) = \alpha_0(t) \, \psi_1 \, \chi_0 +
\beta_0(t) \, \psi_2 \, \chi_0 +
\int d^3 {\bf p} \alpha_{\bf p}(t) \, \psi_1 \, \chi_{\bf p} +
\int d^3 {\bf p} \beta_{\bf p}(t) \, \psi_2 \, \chi_{\bf p}.\nonumber\\
\label{ioniz_of_A_1}
\end{eqnarray}
Here, the initial conditions are given
by $\alpha_0(0) = 1$, $ \beta_0(0) = 0 $ and
$ \alpha_{\bf p}(0) = \beta_{\bf p}(0) = 0$.
The coupling of atom $A$ to both
the subsystem $B$-$B'$ and the laser field involves
bound-continuum transitions which are normally much less
effective than the bound-bound ones. For this reason, we
may assume that the interactions of $A$
with the laser field and the $B$-$B'$-subsystem is weak
and consider ionization of atom $A$ in
the lowest order of perturbation theory in these two interactions.
As a result, by inserting the expansion (\ref{ioniz_of_A_1}) into
the corresponding Schr\"odinger equation we obtain
\begin{eqnarray}
i \frac{ d\alpha_{\bf p} }{dt} - \epsilon^A_p \alpha_{\bf p} & = &
\exp(- i \epsilon^A_g t)  \left\langle \psi_1 \, \chi_{\bf p}
\left|\hat{W}_{A} +
\hat{V}_{AB} + \hat{V}_{AB'} \right| \psi_1 \, \chi_0 \right\rangle
\nonumber \\
i \frac{ d\beta_{\bf p} }{dt} - \epsilon^A_p \beta_{\bf p} & = &
\exp(- i \epsilon^A_g t) \left\langle \psi_2 \, \chi_{\bf p}
\left|\hat{V}_{AB} +
\hat{V}_{AB'} \right| \psi_1 \, \chi_0 \right\rangle,
\label{ioniz_of_A_2}
\end{eqnarray}
where $\epsilon^A_g$ is the energy of the electron
in the initial state $\chi_0$ of atom $A$
and $\epsilon^A_p$ is the electron energy after the emission.
The probability for ionization of the three-atomic system,
as a function of time, then reads
\begin{eqnarray}
P(t) &=& \int d^3{\bf p}
\left( \mid \alpha_{\bf p}(t) \mid^2 +
\mid \beta_{\bf p}(t) \mid^2 \right).
\label{ioniz_of_A_3}
\end{eqnarray}
Note that equations (\ref{ioniz_of_A_2})
are readily solved analytically. However, the resulting
expressions are somewhat lengthy and will not be given here.

\section{ Results and Discussion }

Based on the results obtained in the previous section,
let us now turn to the discussion of some aspects of
photoionization in a system consisting of one lithium
and two helium atoms. We suppose that in our three-atomic system
the positions of the lithium and helium atoms are given by the vectors
${\bf R}_{\rm Li} = (0,0,0)$, ${\bf R}_{\rm He} = (0,0,Z)$
and ${\bf R}_{{\rm He}'} = (0,0,-Z)$, respectively.
Our system is initially (at time $t=0$) in its ground configuration
and is irradiated by a monochromatic laser field.
The field is linearly polarized along the $Z$-axis and its
frequency is resonant to the $\varphi_g$ - $\varphi_+$ transition
in the He-He subsystem, i.e., $E_{+} - E_{g} - \omega_0
= 0$.

Choosing $Z=14$ a.u. we obtain that the energy spitting
$\Delta E_{\pm} = \left| E_{+} - E_{-} \right|$ between
the states $\varphi_{+}$ and $\varphi_{-}$ of the He-He
subsystem is $5.4 \times 10^{-4}$ eV. Assuming a field strength
of $F_0 = 10^{-5}$ a.u., 
the corresponding Rabi frequency amounts to $\Omega_{R_{\rm He-He}} =
2 \left| W_{g,+} \right| = 1.3 \times 10^{-4}$ eV
which is much less than $\Delta E_{\pm} $.

\begin{figure}[t]
\vspace{-0.25cm}
\begin{center}
\includegraphics[width=0.77\textwidth]{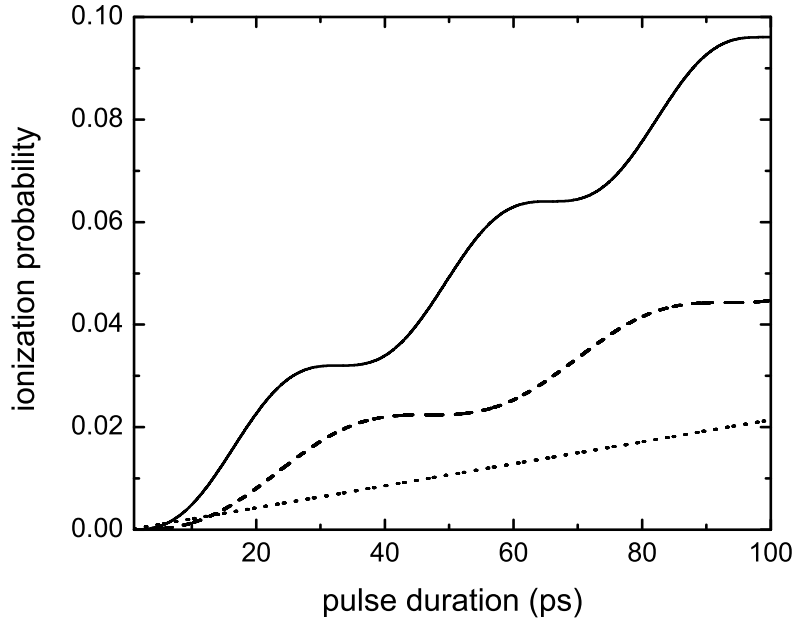}
\end{center}
\vspace{-0.75cm}
\caption{ \footnotesize{
Photoionization probability for Li, Li-He and Li-He-He systems
in an external electromagnetic field, given as a function of time.
The field strength is $F_0=10^{-5}$ a.u.,
the field is linearly polarized and its frequency is resonant to the
corresponding transition in the He or He-He subsystem.
The distances between Li and each of the He atoms is always $14$ a.u.
The atomic positions are aligned along the field polarization with the Li atom
in the middle of the three-atomic system.
The solid, dash and dot curves display results
for Li-He-He, Li-He and Li systems, respectively.
Note that the ionization probability
for an isolated Li atom has been multiplied by a factor of $500$.
For more explanation see the text.} }
\label{time-develop}
\end{figure}

In figure \ref{time-develop}, we present
the probability for ionization
of our system as a function of time.
The probability shows a non-monotonous behaviour
in which time intervals,
when the ionization probability rapidly increases,
are separated by intervals, when the probability
remains practically constant, reflecting oscillations
of the electron populations with the Rabi frequency
$\Omega_{R_{\rm He-He}}$ between
the ground and excited states of the He-He subsystem
in a resonant electromagnetic field.

For comparison, we also show in figure \ref{time-develop} results for
ionization of a single (separated) Li atom and for
ionization in a two-atomic Li-He system.
In the latter case, the lithium atom is located at the origin
(${\bf R}_{\rm Li} = (0,0,0)$) and the coordinates of the helium atom
are ${\bf R}_{\rm He} = (0,0,14\,{\rm a.u.})$. The frequency of the laser field
is assumed to be resonant to the $1s^2\,^1S$--$1s2p\,^1P$ transition frequency of
the corresponding bound states of a single He atom.

In contrast to the single-atom ionization,
in both the two- and three-atomic cases
the ionization probability
demonstrates a step-wise temporal development
in which time intervals of rapid probability growth
are followed by intervals of almost constant probability.
We point out that in
the three-atomic case, however, the size of these
time intervals is shorter by a factor of $\sqrt{2}$.

Compared to ionization of a single Li atom,
ionization in the two-atomic system is very strongly
enhanced \cite{2CPI,ABV1,ABV2}. When the three-atomic system is irradiated, the enhancement increases even further.
In the range of small values of $t$, where
all ionization probabilities still increase monotonously,
this additional enhancement is equal to a factor of $4$.
At larger $t$, however, when the two ionization probabilities
exhibit step-wise behaviours, this additional enhancement 
due to the presence of the second He atom
is reduced to a factor close to $2$ on average,
as can also be seen in figure~\ref{time-develop}.

All the above features can be understood by noting the following:

i) For the chosen set of parameters of our two- and three-center systems,
the indirect channels of ionization, which involve two- or three-atomic
correlations, are substantially stronger than the direct one.
Therefore, these correlations have a dominating effect
on the ionization.

ii) At small $t$, ionization in the two- and three-atomic
systems is basically a two-step process: the first step is
photoexcitation in the He or He-He subsystem and the second step is
a consequent energy transfer to Li. In each case,
both these steps are described by basically the same
dipole transition matrix element of the subsystem.
Since, compared to a single He atom, this dipole element
in He-He is by $\sqrt{2}$ larger than in He,
one obtains a factor of 2 for the enhancement in the ionization amplitude,
leading to a factor of $4$ in the ionization probability (see also \cite{collective}).

iii) At larger $t$, when Rabi oscillations show up, the second step
``saturates'' in the sense that the averaged probability to find
the corresponding subsystem in the excited state becomes
equal to 50\%. Therefore, the ionization probability in
the three-atomic system is now larger (on average)
by a factor of $2$ only.

iv) The origin of the step-wise behaviours
of the ionization probabilities
for the two- and three- atomic systems
lies in the oscillations of
the population between the ground and excited states
in the He atom (for the two-atomic case) or in the
He-He subsystem (for the three-atomic case).
The scale of these oscillation is set
by the Rabi frequency and, because in the
He-He subsystem the latter is
larger by a factor of $\sqrt{2}$,
the corresponding time intervals are shorter
by the same factor.

\begin{figure}[t]
\vspace{-0.25cm}
\begin{center}
\includegraphics[width=0.55\textwidth]{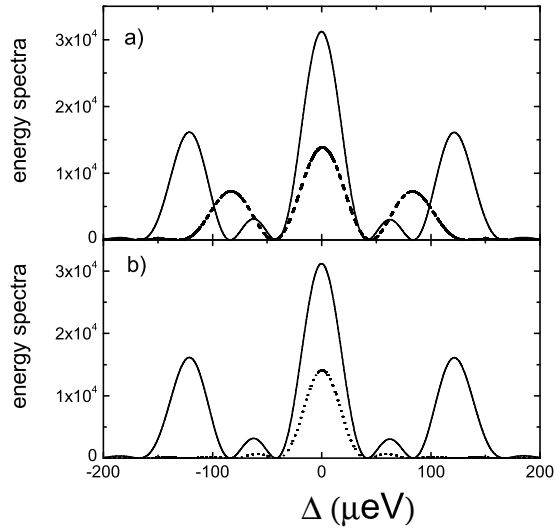}
\end{center}
\vspace{-0.75cm}
\caption{ \footnotesize{
Energy spectra of the emitted electrons, as a function
of $\Delta = \epsilon_p^A - \epsilon_0^A - \omega_0$,
for the same parameters as in figure \ref{time-develop}.
The pulse duration is $100$ ps.
a) Solid and dash curves show results for
ionization of Li-He-He and Li-He systems, respectively.
b) Solid and dot curves display results for
ionization of Li-He-He and Li systems, respectively.
The results for the Li system have been multiplied by a factor of $500$. } }
\label{spectra}
\end{figure}

Additional information about the ionization process
can be obtained by considering the energy spectrum
of emitted electrons.
Such a spectrum is shown in figure \ref{spectra} for
the same systems and parameters as in figure \ref{time-develop} and
for a pulse duration of $T=100$ ps.

In panel (a), we compare the energy spectra of
electrons emitted in the process
of photoionization of Li-He-He and Li-He systems.
In both cases, the main feature is the presence
of three pronounced maxima.
The origin of these peaks is similar to the splitting
into three lines of the energy spectrum of photons
emitted during atomic fluorescence in a resonant
electromagnetic field \cite{mollow1969}.
In such a field, the ground and excited levels of the He and
He-He subsystems split into
two sub-levels, which differ by the corresponding Rabi frequency $\Omega_R$.
As a result, the resonant electronic correlations between these subsystems and the Li atom
lead to an energy transfer to the Li which peaks
at $\omega_0$ and  $\omega_0 \pm \Omega_R/2$.
Since, as was already mentioned, the Rabi frequencies
of these subsystems differ by a factor of $\sqrt{2}$, the magnitude of
the separation between the corresponding
maxima in panel (a) of figure \ref{spectra} also
differs by this factor.
Note also that the widths of these main maxima as well as
the appearance of additional multiple maxima, seen in the figure,
are related to the finiteness of the pulse duration; 
the distance between the latter is roughly given by $2 \pi/T$.

The distinct influence, which the interatomic electron-electron correlations
exert on the shape of the photoelectron spectra, is further highlighted in 
panel (b) of figure \ref{spectra}. It compares the energy spectra of photoelectrons 
emitted from our Li-He-He system and an isolated Li atom. In the latter case,
there is only one main maximum, while the two main side peaks are missing, 
as one would expect (the additional multiple maxima
are related again to the finiteness of the pulse duration).

\section{Conclusion}

We have studied resonant photoionization in a system $A$-$B$-$B'$ consisting of
three atoms, with two atoms $B$ of the same element and one different atom
$A$. We have shown that the mutual correlations among the atoms can largely
enhance the ionization probability and distinctly modify also other
properties of the process in a characteristic manner. In particular, as
compared to the case of resonant photoionization in a two-atom system $A$-$B$,
it has been demonstrated that the presence of a second atom $B$ can (i)
further enhance the photoionization process,
(ii) change the time dependence of the ionization probability and
(iii) move the side peaks in the photoelectron spectrum further apart.

\ack A.B.V. acknowledges the support from the Extreme Matter Institute EMMI.

\end{document}